\documentclass[prl,twocolumn,superscriptaddress,aps]{revtex4-1}
\usepackage{amssymb,graphics,graphicx}
\usepackage{latexsym}
\usepackage{amsfonts}
\usepackage{amsmath}
\usepackage{float}
\newcommand{\be}{\begin{equation}}
\newcommand{\ee}{\end{equation}}
\newcommand{\bea}{\begin{eqnarray}}
\newcommand{\eea}{\end{eqnarray}}
\newcommand{\bma}{\begin{matrix}}
\newcommand{\ema}{\end{matrix}}
\newcommand{\nn}{\nonumber}
\newcommand{\bml}{\begin{mathletters}}
\newcommand{\eml}{\end{mathletters}}
\newcommand{\bes}{\begin{subequations}}
\newcommand{\ees}{\end{subequations}}

\newcommand{\bi}{\begin{itemize}}
\newcommand{\ei}{\end{itemize}}

\begin{document}
\title{Dynamical Electroweak Symmetry Breaking with a heavy fermion in light of recent LHC results}
\author{Pham Q. Hung}
\email{pqh@virginia.edu}
\affiliation{Department of Physics, University of Virginia,
Charlottesville, VA 22904-4714, USA}

\date{\today}

\begin{abstract}
The recent announcement of a discovery of a possible Higgs-like particle -its spin and parity is yet to be determined- at the LHC with
a mass of 126 GeV necessitates a fresh look at the nature of the electroweak symmetry breaking, in particular if this newly-discovered particle
will turn out to have the quantum numbers of a Standard Model Higgs boson.  Even if it were a $0^+$ scalar with the properties expected for a SM Higgs boson,
there is still the quintessential hierarchy problem that one has to deal with and which, by itself, suggests a new physics energy scale around 1 TeV. This article
presents a mini-review of one possible scenario: The formation of a fermion-antifermion condensate coming from a very heavy fourth generation and carrying
the quantum number of the the SM Higgs field and thus breaking the electroweak symmetry. 

\end{abstract}

\maketitle
\section{Introduction}
The nature of the electroweak symmetry breaking is one of the most (if not the most) important problems in particle physics for it is at the heart of the mass problem.
The recent discovery of a new boson with a mass $\sim$ 126 GeV strongly suggests that it might be the long-sought-after Higgs boson of the Standard Model. Much remains to be done to firmly establish first the spin and parity of this new particle and second its detailed decay modes although there is now evidence that the 126-GeV object is a SM-like $0^+$ scalar \cite{Aad:2013xqa} . Even if it is firmly established to be a $0^+$ scalar, one would like to know if there are other spin 0 particles lurking around at higher mass scales. A positive answer would suggest  that the scalar sector is richer than the simplest scenario of a minimal SM with one Higgs doublet and three generations of fermions. A somewhat-related question is the nature of the hierarchy problem itself if nothing else- in particular supersymmetry- is found beside the SM Higgs boson \cite{altarelli}. In fact, the discovery of the SM Higgs boson would cry out for a new scale of physics in the TeV region along with perhaps more yet-to-be-discovered particles.

The hierarchy problem would not be present if there were a physical cutoff scale at around 1 TeV. Such a cutoff scale could come from some kind of new strong interactions at around that scale. The quintessential alternative scenario to supersymmetry came under the generic name of Technicolor (TC) \cite{TC}. A whole new gauge group was postulated with the assumption that the corresponding gauge coupling becomes large at a scale of O(TeV) allowing for condensate formation of Technifermions which dynamically breaks the electroweak symmetry. However, the generic TC scenario ran into several serious problems with flavor-changing neutral currents and had to be replaced by a more complicated scenario that goes under the name of Walking Technicolor. The discovery of a new "resonance" at 126 GeV, if confirmed as a Higgs-like boson, puts TC and its extension in difficulty since these scenarios favor heavier spin-0 particles. 

There is another possibility of a strong dynamics arising at a TeV scale \cite{pqchi1,pqchi2,pqchi3}: The formation of EW condensates coming from a heavy fourth generation through the exchange of a massless scalar \cite{holdom}, \cite{pqchi2}. Here we describe the approach of \cite{pqchi2}. This model is scale invariant at the classical level. It was argued in  \cite{pqchi2} that starting from some value of the Yukawa couplings at the electroweak scale which corresponds to a heavy fourth generation, these couplings reach the critical value for condensate formation at an energy around O(TeV). Those condensates are bound states of a fourth generation fermion and anti-fermion, very much like the TC scenario but, at the same time, very much unlike it because it does not require a new strong interaction gauge group. Dynamical electroweak symmetry breaking can be achieved with the simplest extension of the SM: the postulated existence of a heavy fourth generation. This review article will describe in steps the conditions for condensate formation, the scale where this condensation occurs, the nature of the Higgs scalars and the possible connection to the 126 GeV "resonance" along with some phenomenology, and, last but not least, a summary of the extension of the SM with four generations to a model which is conformally invariant above O(TeV).

\section{Bound states and condensates of heavy fourth generation fermions}

Bound states of a fourth generation fermion and anti fermion can be formed when the Yukawa couplings to the Higgs field is sufficiently large. There are bound states that are loose and there are those that are so strongly bound that they become condensates which can break the electroweak symmetry. We first begin with a discussion of bound state formation using a simple-minded non-relativistic approach. This discussion sets the stage for the second part which will discuss condensate formation in the following sense: Under what conditions can a fermion-antifermion bound state get formed in this simple-minded approach and what are the fermions which can do so? It was shown in \cite{pqchi1} that, among the heavy quarks, only the 4th generation quarks which will be assumed to be heavier than the top quark can form bound states. It should be clarified at this point that this simple-minded non-relativistic approach which is based on a scenario in which the Higgs field is self-interacting with a non-vanishing mass. It was indicated in \cite{pqchi1} and shown below that these states become more tightly bound as the mass and self-coupling tend to zero. In this limit of vanishing mass and self-coupling, the non-relativistic approach breaks down. 

We present next a fully relativistic approach using the Schwinger-Dyson  (SD) equation \cite{pqchi2} to investigate the conditions for condensate formation by writing down an integral equation for the fermion self energies under the rainbow approximation. Based on the aforementioned observation, we started out with a scale-invariant Higgs-Yukawa system in which the scalar field is non-self interacting \cite{pqchi2}. Beside the wish to have dynamical electroweak symmetry breaking coming from condensates formed by the exchange between a fermion and antifermion  of a massless,  non-self interacting scalar, there is a possibility of embedding this scale-invariant model in a larger framework which possesses conformal symmetry \cite{chiumanpqtom}.

At this point, a clarification is in order before we begin our discussion of bound state formation. As we will discuss below, the consequences of the bound state dynamics presented here are the existence of three doublets of Higgs fields: Two composite doublets and the original doublet. Although, in the construction of bound states and condensates, the fundamental Higgs scalar which is used as a force carrier and couples to all relevant fermions with the corresponding Yukawa couplings, the 126 GeV boson, one of the mass eigenstates of the three Higgs doublets, might couple  differently to different fermions and  in particular its coupling to the fourth generation quarks might be suppressed. This depends on the detailed mixings among the three Higgs doublets. 

\subsection{Bound states through Higgs exchange}

The simplest approach which shows the essence of the formation of a bound state of a heavy 4th-generation fermion with an anti-fermion  is through the use of the non-relativistic Schr\"{o}dinger equation, neglecting spin and relativistic corrections. In what follows we describe the analysis performed by \cite{pqchi1}. A more complete analysis taking into account the spin of the fermions as well as relativistic corrections to the Yukawa potential can be found in \cite{wise}. 

\begin{figure}[H]
\centering
    \includegraphics[scale=0.77]{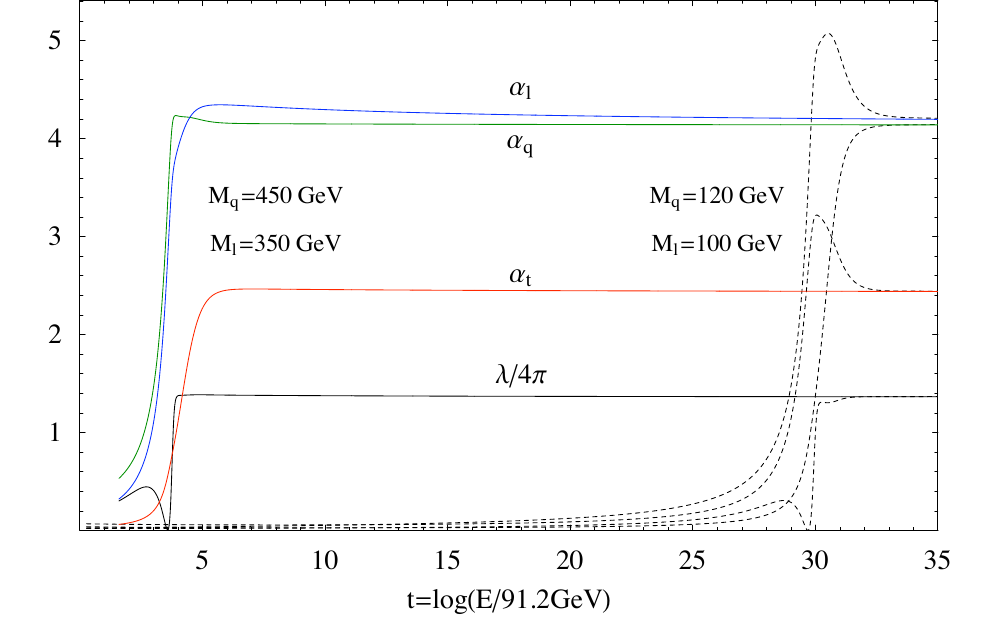}
\caption{{\small The evolution at 2 loops of $\alpha_f = g_{f}^2/4\pi$ where $g_f$ is the Yukawa coupling for the quarks and leptons of the 4th generation and for the top quark. Two sets of values are shown with the "lighter" shown solely for comparison}}
\label{RGevolution}
\end{figure}

We concentrate on the case of the exchange scalar being massive and study the behavior of the bound states as the mass approaches zero.
The non-relativistic Higgs-exchange Yukawa potential can be read as follows
\begin{equation}
\label{yukpot}
V(r) = - \alpha_Y(r) \frac{e^{-m_H(r) r}} {r},
\end{equation}
where $m_H$ is the Higgs mass and $\alpha_Y = \frac{m_1 m_2 } {4\pi v^2} $ with $v=246$ GeV. The masses of the two "fermions" are $m_1 $ and $m_2$ and the reduced masses of the system is $ M = m_1 m_2 /(m_1 +m_2)$. Ref \cite{pqchi1} used the Rayleigh-Ritz variational method to gain some insight into constraints from bound state formation and followed up with a more accurate numerical solution to the Schr\"{o}dinger equation. A trial wave function, $u (y, r)=2 y^{\frac{3}{2}}~e^{-y r}$, was used where $y$ is the variational parameter. The condition for bound state formation can be obtained by looking for the optimum relative  energy
\begin{equation}
\label{openergy}
E=-\alpha_Y m_H \frac{z^3(z-1)}{4 (z+1)^3},
\end{equation}
where $ z=2y/m_H$ is a redefined variable. For simplicity,  \cite{pqchi1} specialized to the case where $m_1=m_2=m_f$ and introduced a function $K_f \equiv \alpha_Y m_f/m_H = (1+z)^3/z(z+3)$. The condition for bound state formation (negative relative energy) follows from Eq. \ref{openergy}, namely $z>1$ giving
\begin{equation}
\label{condition0}
K_f = \alpha_Y m_f/m_H = \frac{g_f^3}{16 \pi \sqrt{\lambda}}>2 ,
\end{equation}
where $m_f = g_f (v/\sqrt{2})$ and $m_H =\sqrt{2 \lambda} v$ have been used. As one can see from (\ref{condition0}), the condition for bound state formation depends only on the relative strengths of the Yukawa couplings and the Higgs quartic coupling. The lower bound in (\ref{condition}) gets modified when the Schr\"{o}dinger equation is solved numerically 
\begin{equation}
\label{condition2}
K_f = \alpha_Y m_f/m_H = \frac{g_f^3}{16 \pi \sqrt{\lambda}}> K_0=1.68 .
\end{equation}
To appreciate the bound (\ref{condition2}), \cite{pqchi1} studied the evolution of the couplings at two loops for the quartic Higgs self coupling, the 4th family and the top quarks Yukawa couplings (all others were too small to play any role). An example is shown in FIG.~\ref{RGevolution}.
The values of $K_f$ for the 4th family quark and leptons and for the top quark evaluated at a scale O(TeV) was found to be $K_q=1.82$, $K_l=1.92$ (corresponding to masses $m_q= 450\, GeV$ and $m_l=350\,GeV$), and $K_t=0.82$. Comparing these numbers with (\ref{condition2}) we can see that the 4th generation quarks and leptons form rather loose bound states (in the sense that the binding energy is small) while the top quark is too light to form bound states through a Higgs exchange. A more detailed investigation of this kind of bound states was carried out by (\cite{wise}) which include corrections taking into account spin and relativistic effects among others. The binding energies calculated in \cite{wise} are consistent with those obtained using the aforementioned simple approach. 

\begin{figure}[H]
\centering
    \includegraphics[scale=0.77]{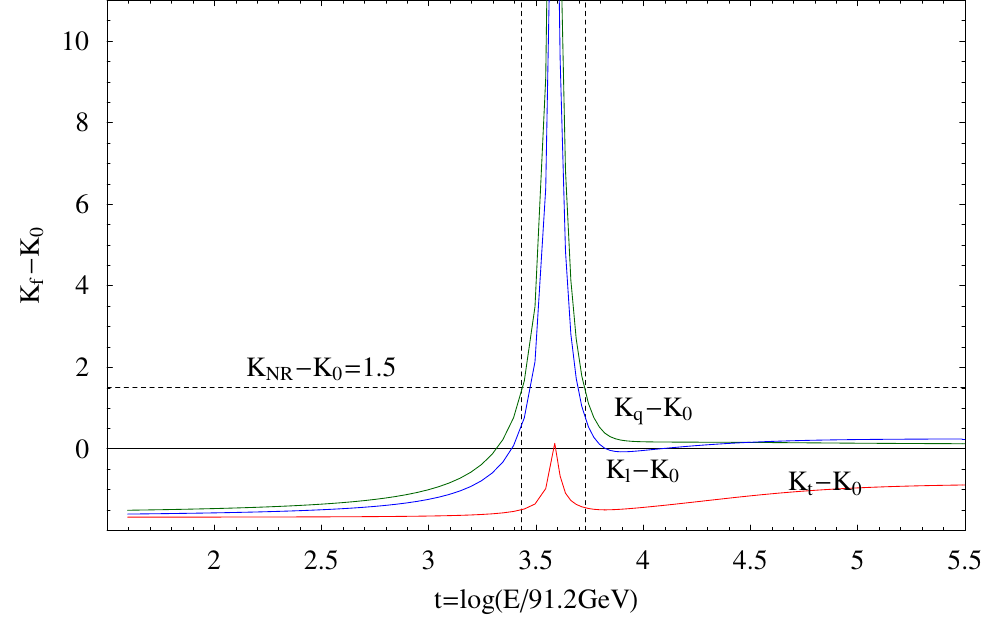}
\caption{{\small ($m_q=450$ GeV and $m_l=350$ GeV)~$K_f - K_0$ with $ K_f = g_f^3 / 16 \pi \sqrt{\lambda}$ and $K_0=1.68$. For illustration purpose, the initial value of $\lambda$ is increased slightly such that the peak value of $K_f$ would not become too large to fit in the figure. The horizontal dotted line indicates an estimate of $K_f$ where the non-relativistic method is still applicable and the vertical dotted lines enclose the region where a fully relativistic approach is needed.}}
\label{condensates}
\end{figure}

As the Higgs quartic coupling, $\lambda$, decreases ($m_H$ decreases), one would expect from (\ref{condition2}) that the bound states will become more tightly bound until condensate formation becomes possible. However, relativistic effects become more and more important as the Higgs mass becomes smaller and smaller. This is shown heuristically in FIG.~\ref{condensates} (\cite{pqchi1}).
In FIG.~ \ref{condensates}, the rise in value for $K_f - K_0$ corresponds to the decrease in the Higgs mass. Furthermore, if one looks back at FIG.~\ref{RGevolution}, one notices that there is a "dip" in the Higgs quartic coupling where it vanishes before rising again. The Yukawa potential becomes more and more Coulomb-like. At the dip, one has, with $\alpha_Y \sim 1.6$,
\begin{equation}
\label{dip}
V_{"dip"}(r) = - \frac{\alpha_Y} {r},
\end{equation}
Eq.~(\ref{dip}) represents a strong attractive Coulomb-like potential. It was mentioned in \cite{pqchi1} that such a potential has been studied in condensed matter physics where it was found that it could potentially lead to the formation of condensates \cite{CDM}. 

As it has been mentioned at the beginning of this section, in this region where a condensate could potentially get formed, the non-relativistic approximation breaks down and a fully relativistic approach is needed. This is also a region of vanishing scalar mass and self coupling.
It is then natural to ask, if condensates were to get formed and under what conditions would that be so, if we start out with a massless scalar at tree level. In particular, we assume that the model is scale invariant and that the fundamental Higgs doublet does not obtain a non-vanishing vacuum expectation value (VEV) at tree level. If condensates carrying the quantum numbers of the SM can get formed via the exchange of the massless scalar, scale invariance will be broken spontaneously. The breaking of scale invariance and of the electroweak symmetry seem to be deeply related in this scenario.

\subsection{Condensates through massless Higgs exchange}

In this section, we present an analysis of condensate formation of a heavy fermion anti-fermion system by the exchange of a massless Higgs scalar carried out by \cite{pqchi2}. Basically, the solution sought in \cite{pqchi2} was based on the SD equation in the ladder (or rainbow) approximation to the contribution to the self-energy of the heavy fermion. The assumptions made in \cite{pqchi2} are basically as follows: 1) It is assumed that the fundamental Higgs doublet has no mass term and cannot develop a VEV at tree level; 2) Only the Yukawa couplings of the 4th generation are important enough to participate in this process; 3) All gauge interactions will be neglected. The Yukawa interaction Lagrangian for the 4th generation quarks can be written as
\begin{equation} \label{action}
\mathcal{L}_Y = - g_{b'} ~\bar{q}_L \Phi ~b'_R - g_{t'} ~\bar{q}_L \widetilde{\Phi} ~t'_R + h.c. \,,
\end{equation}
where $\widetilde{\Phi}= i \tau_2 \Phi^{*}$ and $q_L =(t', b')_L$ . A similar Lagrangian can be written for the 4th generation leptons. The steps that we used in \cite{pqchi2} are as follows: 1) We first calculate the self-energy $\Sigma_{4Q, 4L}(p)$ of the 4th generation fermions and find the condition for condensate formation; 2) Next we calculate the corresponding condensates $\langle \bar{t'}_L t'_R \rangle$,$\langle \bar{b'}_L b'_R \rangle$ as well as those for the 4th generation leptons. The integral equation for the fermion self-energy has been calculated in two different ways resulting in the same result. Here we present a diagrammatic version of the calculation for clarity. We wish to stress that this section is different in nature from the preceding one in that we {\em do not make use of non-relativistic concepts} used in that section but simply compute the self-energy of the fermion.

The diagrams which contribute to the SD equation in the ladder approximation coming from the exchanges of the {\em massless complex} $\phi^0$ and $\phi^+$ fields are illustrated below (see the note \cite{note}):
\begin{figure}[H]
\includegraphics[angle=0,width=8.5cm]{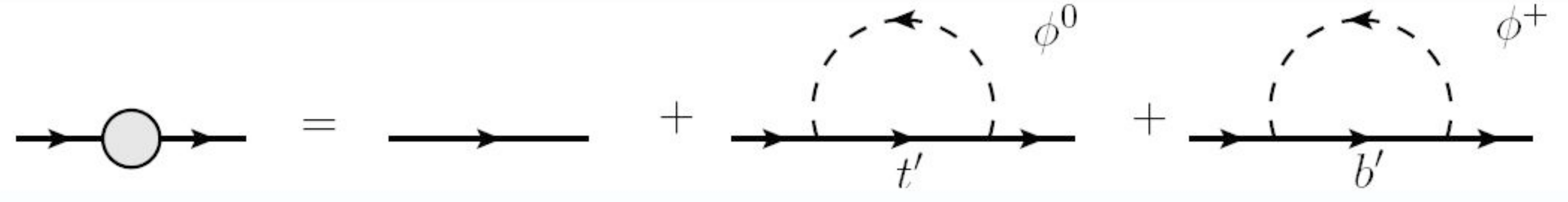}
\caption{{\small Graphs contributing to the right-hand side of the Schwinger-Dyson equation for the fermion self-energy $\Sigma(p)$ for $t'$. Similar graphs for the $b'$ self-energy are obtained by the substitution $t' \leftrightarrow b'$. The self-energies of the 4th leptons are computed in the same way.}}
\label{SDEfig}
\end{figure}
The SD equation can then be written as (for the 4th generation quarks)
\begin{equation} \label{SD}
\Sigma_{4Q}(p) = \frac{+2 g_{4Q}^2}{(2 \pi)^4}\int d^4 q \frac{1}{(p-q)^2} \frac{\Sigma_{4Q}(q)}{q^2 + \Sigma_{4Q}^2(q)} \,.
\end{equation}
Notice that custodial symmetry is imposed by making $g_{t'}=g_{b'}$ resulting in the factor of 2 on the right-hand side of Eq.~(\ref{SD}) coming from the equality of the last two diagrams in FIG.~\ref{SDEfig}.

This equation can be transformed into a differential equation
\begin{equation}
\label{SDdiff}
\Box \Sigma_{4Q}(p) = - (\frac{\alpha_{4Q}}{\alpha_c}) \frac{\Sigma_{4Q}(q)}{q^2 + \Sigma_{4Q}^2(q)} \,.
\end{equation}
with the boundary conditions 
\bea \label{bc1}
\nn
&&\lim_{p \to 0} p^4 \frac{d \Sigma_{4Q}}{d p^2} = 0 \,, \\
&&\lim_{p \to \Lambda}  p^2 \frac{d \Sigma_{4Q}}{d p^2} + \Sigma_{4Q}(p)  = 0 \,.
\eea

In (\ref{SDdiff}), $\alpha_c$ is the {\em critical coupling} given by
 \begin{equation} \label{crit}
\alpha_c = \pi/2 .
\end{equation}
This critical coupling determines the condition in which condensates can get formed as we shall see below.
A change of variables
\bea 
\nn
p &=& ~e^{t} \\
\Sigma_{4Q} (p) &=& e^t ~u(t-t_0)  \,,
\eea
transforms Eq. (\ref{SDdiff}) into
\be \label{diff}
\frac{d^2 u}{d t^2} + 4 \frac{du}{dt} + 3 u + (\frac{\alpha_{4Q}}{\alpha_c}) \frac{u}{1+u^2} = 0 \,.
\ee
with the boundary conditions
\bea \label{bc2}
\nn
&&\lim_{t \to t_{\Lambda}} (u' + 3 u) = 0 \,, \\
&&\lim_{t \to -\infty} ( u' + u ) = 0 \,,
\eea

A few remarks are in order.
\bi
\item
As with earlier studies of strong QED dynamics \cite{kugo,bardeen}, it was found that for $\alpha_{4Q} < \alpha_c=\pi/2$, there is no solution except for the trivial one, namely $\Sigma_{4Q} (p) =0$. There is no dynamical symmetry breaking in that case. 
\item
For $\alpha_{4Q} > \alpha_c=\pi/2$, a nontrivial solution for $\Sigma_{4Q} (p)$ is found \cite{pqchi2} and can be written as
\be \label{sigmasol}
\Sigma_{4Q} (p) \sim p ^{- 1} \sin [\sqrt{\frac{\alpha_{4Q}}{\alpha_c} -1} (\ln p + \delta)] \,.
\ee
The dynamical mass of the 4th generation quarks corresponding to the vacuum solution with the largest fermion self-energy is given by
\be \label{sigma0} 
\Sigma_{4Q} (0) = \Lambda e^{1 - \pi/\sqrt{\frac{\alpha}{\alpha_c} -1} + \delta_0}~,
\ee
In (\ref{sigma0}), $\delta_0$ is the phase which corresponds to the vacuum solution. $\Lambda$ is the physical cutoff scale where dynamical symmetry breaking takes place. Notice that the total mass of a 4th generation quark, e.g. that of $t'$, is $m_{t'} = \Sigma_{4Q}(0)$+ Lagrangian mass.
\item
The quantities that are most relevant to dynamical electroweak symmetry breaking are the condensates of 4th generation fermions which carry the electroweak quantum numbers of a Higgs doublet. They can be computed once we know the self-energy $\Sigma_{4Q} (p)$. 
\bea \label{tbart1}
\nn
\langle \bar{t'}_L t'_R \rangle=\langle \bar{b'}_L b'_R \rangle  &=& - \frac{3}{4 \pi^4} \int d^4 q \frac{\Sigma_{4Q}(q)}{q^2 + \Sigma_{4Q}^2(q)} \\
                            &=& \frac{3}{2 \pi^2} (\frac{\alpha_c}{\alpha})~e^{3 t_{\Lambda}} [ u'(t_{\Lambda} -t_0) + u(t_{\Lambda} -t_0)] \nonumber \\
                            &=& \frac{-3}{\pi^2} (\frac{\alpha_c}{\alpha})~e^{3 t_{\Lambda}} u(t_{\Lambda} -t_0) \nonumber \\
                           & \approx&- \frac{3}{\pi^2} (\frac{\alpha_c}{\alpha_{4Q}})~\Lambda~\Sigma_{4Q}^2(0)~
\sin[\sqrt{\frac{\alpha_{4Q}}{\alpha_c}-1}] \nonumber \,,\\  
 \eea
 where the Yukawa couplings for $t'$ and $b'$ are chosen so as to guarantee an $SU(2)$ custodial symmetry. We require 
 \be \label{condition}
\langle \bar{t'}_L t'_R \rangle \sim O(-\Lambda_{EW}^3)\,.
\ee
\item
From (\ref{tbart1}) and (\ref{condition}), we observe that the physical cutoff scale $\Lambda$ plays an important role in determining whether or not one needs to fine tune the theory in order to fit experiment. Since $\Sigma_{4Q}^2(0) \sim O(\Lambda_{EW}^2)$, it is straightforward to see that, for $\Lambda \sim O(TeV)$, $\alpha_{4Q}$ does not need to be close in value to $\alpha_c$ in order  to satisfy (\ref{condition}).  Hence, no fine tuning! For $\Lambda \sim 10^{16} GeV$, one needs to fine tune $\alpha_{4Q}$ down to 28 decimal points i.e. $\alpha_{4Q}/\alpha_{c} - 1 \sim 10^{-28}$!
\item 
Condensates of the 4th generation leptons can be obtained in a similar manner \cite{pqchi2}. One has
\be \label{LbarL}
\langle \bar{L}_L L_R \rangle=  \langle \bar{N}_L N_R \rangle \approx - \frac{1}{\pi^2} (\frac{\alpha_c}{\alpha_{4L}})~\Lambda~\Sigma_{4L}^2(0)~
\sin[\sqrt{\frac{\alpha_{4L}}{\alpha_c}-1}] \,.
\ee 
The condition for condensate formation is the same as that for the quarks, namely $\alpha_{4L} > \alpha_c =\pi/2$.
\item
To see at around what energy scale $\alpha_{4Q}$ exceeds the critical coupling for condensate formation, it is instructive to evolve the couplings of the Higgs-Yukawa system at one and two loops. This was carried out by \cite{pqchi1,pqchi2}. 
\item
It was argued in \cite{pqchi2} that the fundamental massless Higgs doublet obtains an {\em induced negative mass squared} which would give rise to its own VEV which was absent before condensation occurs.
\be \label{dim5vev}
\frac{1}{2} \{\frac{2 g_{4Q}^2}{ \Sigma_{4Q}(0)} \langle \bar{t'}_L t'_R \rangle + \frac{2 g_{4L}^2}{ \Sigma_{4L}(0)}\langle \bar{L}_L L_R \rangle + H.c. \}|\phi^0|^2 \,.
\ee
 
\ei

\cite{george} has also investigated this aspect of a heavy four generation model from a slightly different point of view. For a more detailed exposure of the ideas used in \cite{george}, one can consult a companion article by George Hou in this special issue.

\section{Renormalization group analysis of the Higgs-Yukawa system for the 4th family}
   
Since the couplings of the Higgs-Yukawa system evolve with energy, it is worthwhile to analyze such a behavior for for the 4th generation. In \cite{pqchi1}, a two-loop renormalization group analysis was carried out. for the 4th generation Yukawa couplings and for the Higgs quartic coupling. It was found that, above some energy scale, a quasi-fixed point for the couplings was observed. This quasi-fixed point, if it is real and is not just an artifact of a two-loop approximation, may point to an interesting domain where conformal invariance may be effective.  In this section, we are mainly interested in the energy scale where the Yukawa couplings exceed the critical coupling $\alpha_c=\pi/2 \approx 1.57$ for condensate formation. In addition, a one-loop analysis was also done in order to compare the energy where the Landau pole of the one-loop approximation appears with that of the two-loop quasi-fixed point.
 \begin{figure}
\includegraphics[angle=0,width=8.5cm]{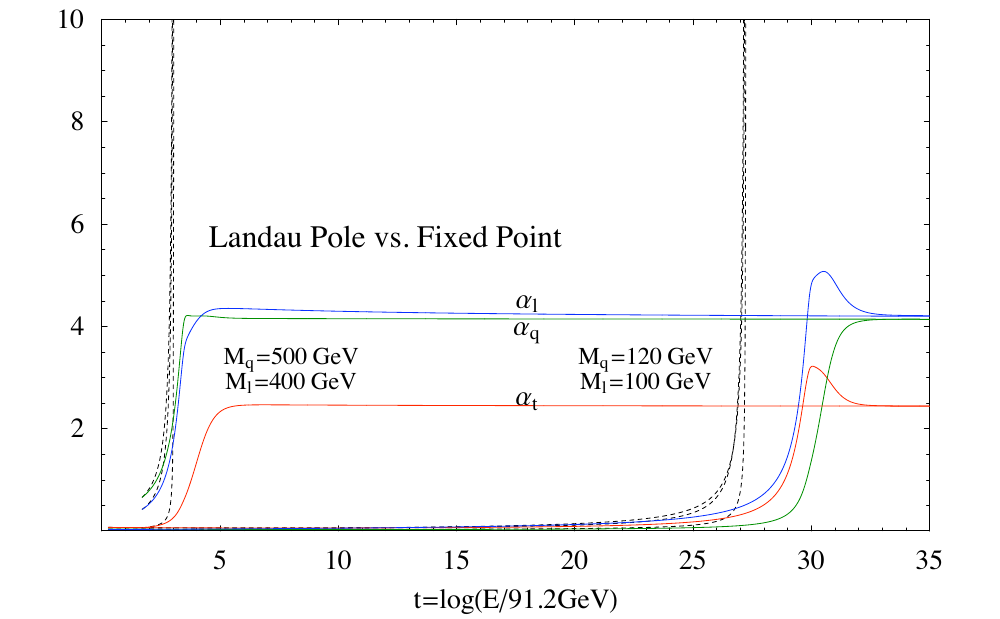}
\caption{{\small The Landau pole(dotted lines) and the quasi fixed point(solid lines) of the Yukawa couplings of the
fourth generation fermions and the top quark. For a heavy fourth generation (left side), both the Landau singularity from
one-loop RGEs  and the quasi fixed point from two-loop RGEs appear at about 2 $\sim$ 3 TeV, while for a light fourth 
generation (right side), their locations at the energy scale differ by two orders of magnitude.  
}}
\label{fig2}
\end{figure}
From (\ref{fig2}), we can see that, for a heavy 4th generation, the one and two-loop approximation more or less coincide with each other until the Yukawa couplings exceed the critical value of 1.57, and this happens at the energy scale 2~3 TeV. In terms of dynamical electroweak symmetry breaking, this scale which represents a physical cutoff scale $\Lambda$ of O(TeV) appears genuine beyond the one and two-loop approximations used. Although it is not shown in (\ref{fig2}), one can easily see that a 600 GeV (at least for the quarks) 4th generation would induce dynamical electroweak symmetry breaking at at scale of O(TeV).

As we have stated at the beginning of this review, a physical cutoff scale of O(TeV) would circumvent the hierarchy problem \cite{pqchi2,pqchi3} although we have started with a Lagrangian which contains a fundamental scalar. The difference with the usual approach is the fact that the assumption of scale invariance forbids a "mass term" for the fundamental Higgs doublet and prevents it from acquiring a VEV at tree level. As it is suggested in \cite{pqchi2}, its induced VEV is proportional to the condensate value and it would naturally be of O($\Lambda_{EW}$).

What would be the implication of a fixed point be if it exists above the scale of dynamical electroweak symmetry breaking as shown below?
\begin{figure}
\includegraphics[angle=0,width=8.5cm]{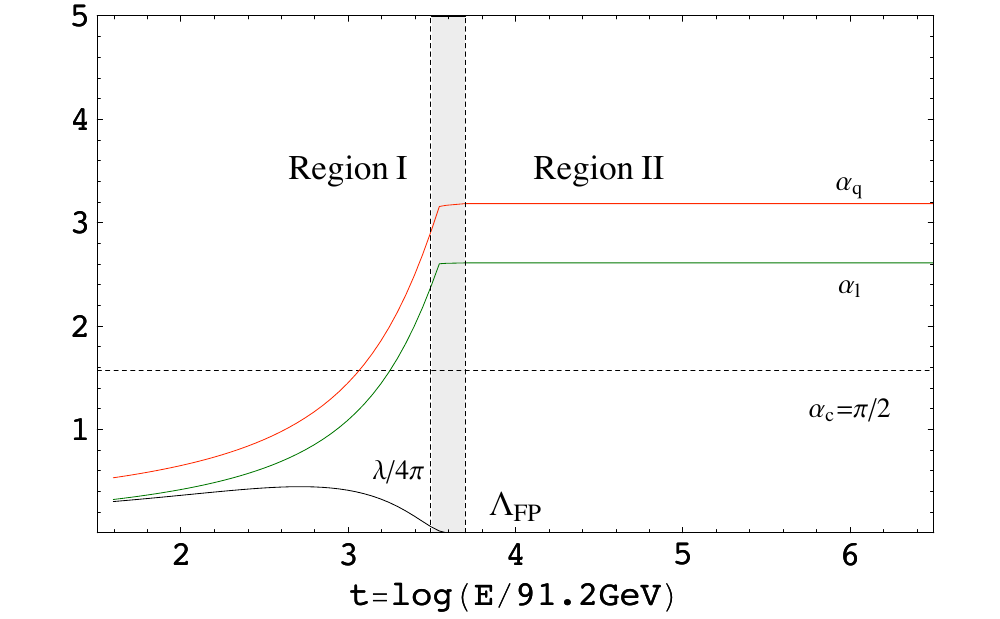}
\caption{A combination of RGE and Schwinger-Dyson analysis}
\label{ideal}
\end{figure}
Here Region II represents a scale-invariant (or even conformally-invariant) energy region while Region I is one in which dynamical electroweak symmetry breaking occurs. Notice that in region II, the Yukawa couplings are constant but the boundary condition of the SD equation is satisfied only at $\Lambda_{FP}$ and {\em that is the symmetry breaking scale}.
Before speculating further on this scenario, let us say a few words about the scalar sector.

\section{The scalar sector of BSM4, FCNC, 4th neutrino, Leptoquarks}

The strong dynamics of this model of BSM4 (BSM4 stands for an extended model of the SM with four generations and more than one scalar doublet) gives rise to three Higgs doublets: one fundamental and two composite doublets. \cite{pqchi2} denotes them as
\bea \label{3higgs}
\nn
H_1 &=& ( \pi^+,\pi^-, \pi^0, \sigma ) \,, \\
\nn
H_2 &=& ( \bar{b'} t', \bar{t'} b', \bar{t'} t'-\bar{b'} b', \bar{t'} t'+ \bar{b'} b' ) \,, \\
H_3 &=& ( \bar{E } N, \bar{N} E, \bar{N} N -
\bar{E} E, \bar{N} N + \bar{E} E ) \,.
\eea
As it is obvious from the notations, $H_2$ and $H_3$ are composite while $H_1$ is the original fundamental doublet. The diagonalization of the scalar mass matrices will yield three NG bosons which are absorbed by W's and Z bosons leaving 6 {\em massive pseudo NG bosons} and 3 {\em massive "Higgs"} scalars. A more precise and detailed statement on the mass spectrum of these scalars are under investigation. Which one might correspond to the observed 126 GeV Higgs-like boson is a question we would like to have an answer to in our model. 

The production cross section of that 126-GeV object appears to exclude 4th generation quarks with masses less than $\sim$ 600 GeV if they couple to the SM Higgs boson in the same manner as the lighter ones. However, it is perhaps premature to "rule out" the 4th generation below 600 GeV. In particular, one would like to know whether or not the coupling of the 4th generation to the 126-GeV object is suppressed and by how much. 

It is well-known that a three Higgs doublet model is notoriously difficult to analyze and various assumptions have to be made in order to reduce the number of parameters in the Higgs potential. Whether the lightest scalar (assumed to have a mass of say 126 GeV) is a $0^+$ or $0^-$ particle is a model-dependent issue. However, recent analyses \cite{Aad:2013xqa} excluded $0^-$, $1^{\pm}$ and $2^+$ at confidence levels above 97.8 \%, leaving $0^+$ as a favored candidate. This puts a severe constraint on any multi-Higgs model. In the present model, the three $0^+$ scalars come from $H_1$, $H_2$ and $H_3$. An exhaustive study of a three-Higgs-doublet model is beyond the scope of this review although a detailed study of the various assumptions is obviously needed. 

However, it is important to illustrate the type of hybrid models in which one of the Higgs doublets is fundamental while the other ones are composite. This is nicely exposed in a companion article \cite{Geller:2013dla}. This is a model with two Higgs doublets: one fundamental and one composite. The fundamental doublet couples mostly to light fermions while the composite one couples only to the 4th generation. In this model, condensates of the 4th generation quarks come from a Nambu-Jona-Lasinio term in the Lagrangian. (The couplings of the 4th generation leptons are assumed to be subcritical such that no condensates are formed in this sector.)  Although, on the surface the model of \cite{Geller:2013dla} differs from that of \cite{pqchi2}, the hybrid nature of both models makes the two models similar in many ways. Below the compositeness scale $\Lambda$, a two-Higgs-doublet potential can be written down and minimized \cite{Geller:2013dla}, taking into account the compositeness conditions. The result of the analysis  \cite{Geller:2013dla} shows that, for a range of mixing angle $\tan \beta = v_h / v_l \leq 0.7$, with $v_h$ and $v_l$ being the VEV of the composite scalar and the fundamental scalar respectively, one can find the lightest $0^+$ scalar to have a mass $\sim$ 126 GeV with the state $0^-$ being heavier and that this $0^+$ state is mostly composed of the fundamental field. The other $0^+$ is much heavier. What the hybrid two-Higgs doublet model of  \cite{Geller:2013dla} shows is the fact that it is possible to find parameters that can give rise to a "light" state (126 GeV) which can mimic the SM Higgs boson. For the hybrid model of \cite{pqchi2} with one fundamental and two composite Higgs doublets, it goes without saying that, with an increase in the number of parameters, it would not be difficult to find a set of parameters that can satisfy the present constraints. However, much work remains to be done to show explicitly the various possibilities. 

The couplings of the lighter three generations to the aforementioned nine massive states will come from the basic couplings to the various components of $H_1$, $H_2$ and $H_3$. In particular, the lighter three generations can couple to $H_2$ (for the quarks) and $H_3$ (for the leptons) only through a one-loop process and the strengths of the couplings are quite constrained by the mixings between the 4th generation and the lighter three in the Yukawa sector. Flavor-changing neutral current (FCNC) processes can be quite suppressed if the 4th generation does not mix much with e.g. the first and second generation.  These points were discussed in sufficient details in \cite{pqchi2}. 

Last but not least is the question of the 4th neutrino. In this scenario, it is easy to see that the 4th neutrino can be quite heavy since it participates in the condensation process and its (Dirac) mass contains the self-energy $\Sigma_{4N}(0)$ which is large. For this simple reason, it is {\em natural} why the 4th neutrino can be much heavier than the other three and cannot contribute to the Z width. However, the subject of neutrino masses is quite vast and is an interesting topic on its own which deserves a close scrutiny.

At this point, it is worth mentioning a further difference between the two models of  \cite{Geller:2013dla} and \cite{pqchi2}. As it can be seen in the discussion of the previous two sections that both 4th generation quarks and leptons can form condensates. This is the reason why in the model of \cite{pqchi2} there are {\em two} composite Higgs doublets $H_2$ and $H_3$ formed from quarks and leptons of the 4th generation respectively (Eq.~\ref{3higgs}). Because of this, one may expect bound states between a 4th generation quark and a 4th generation lepton. These bound states of quarks and leptons would form the so-called "leptoquarks". These are "meson"-like particles of varying spins (0,1,..)  and can be very heavy. With $Q_L$ and $L_L$ being the 4th generation quark and lepton doublets respectively, one has for example the following scalar leptoquarks:
\be
\label{lepto1}
\bar{Q}_{L}\, E_R = \left( \begin{array}{c}
\bar{t}^{'}_L \, E_R \\
\bar{b}^{'}_L \, E_R ,
\end{array} \right)
\ee
with charges $-5/3,-4/3$;
\be
\label{lepto2}
\bar{Q}_{L}\, N_R = \left( \begin{array}{c}
\bar{t}^{'}_L \, N_R \\
\bar{b}^{'}_L \, N_R ,
\end{array} \right)
\ee
with charges $-2/3,1/3$. 
\be
\label{lepto3}
\bar{L}_{L}\, t^{'}_R = \left( \begin{array}{c}
\bar{N}^{'}_L \, t^{'}_R \\
\bar{E}^{'}_L \, t^{'}_R ,
\end{array} \right)
\ee
with charges $2/3, 5/3$.
\be
\label{lepto3}
\bar{L}_{L}\, b^{'}_R = \left( \begin{array}{c}
\bar{N}^{'}_L \, b^{'}_R \\
\bar{E}^{'}_L \, b^{'}_R ,
\end{array} \right)
\ee
with charges $-1/3,2/3$.
Higher spin leptoquarks can in principle exist. These can leave spectacular signatures. It is beyond the scope of this review to go into details on the formation, production and decay of these leptoquarks. This discussion will be presented in  \cite{hunglepto}.

\section{Epilog}

A heavy fourth generation is not only a viable scenario \cite{fourth} but is also an attractive possibility for breaking the electroweak symmetry dynamically through condensations of 4th generation fermions. It also provides a natural setting for the existence of a physical cutoff scale of O(TeV) without having to go outside the existing gauge structure of the SM: The dynamics for condensate formation is provided by the exchange of a massless fundamental Higgs scalar between 4th generation fermions. In a certain way, BSM4 contains its own "solution" to the hierarchy problem. The discovery of the 126-GeV "object", if confirmed as a $0^{+}$ particle, cries out for more particles since its sole existence would still leave a big question mark for the SM, that of the hierarchy problem.

BSM4 as discussed in this review has a rich scalar spectrum: Two composite Higgs doublets and one fundamental. An interesting phenomenology lies ahead in the search for extra particles beyond the 126-GeV object. Since the 4th generation quarks can be quite heavy to be directly probed in the near future, they might manifest themselves through characteristics of extra spin-0 particles (if found).  There could also be composite leptoquarks formed from a bound state of a quark and a lepton of the 4th generation \cite{hunglepto}.Other interesting aspects of BSM4 are discussed in companion articles of this special issue.  

We end with some speculation concerning what might lie beyond the scale where dynamical electroweak symmetry breaking occurs. What \cite{pqchi1} has shown within a two-loop approximation was the existence of a quasi-fixed point meaning that $\beta_{2loop} \approx 0$ beyond $\Lambda_{cutoff}$ of O(TeV). This "ultraviolet" quasi-fixed point might point to a regime where conformal invariance is unbroken. It is within this context that \cite{chiumanpqtom} has constructed non-SUSY conformal four family models from orbifolded $AdS_{5} \otimes S^5$. It was found, in this construction, that three of the four families come from the same representation while the fourth one comes from a different initial representation. To quote from  \cite{chiumanpqtom}: This appears to single out the fourth generation as somewhat different from the other three. New gauge bosons, fermions and scalars are predicted in this model.

\section{Acknowledgments}

I would like to thank George Hou and the editors of this special issue for the invitation to write a review on dynamical electroweak symmetry breaking with SM4. This work was supported by US DOE grant DE-FG02-97ER41027.


\begin{thebibliography}{99}
\bibitem{note} For pedagogical purpose, it should be noted that the propagator of a complex scalar field is $\langle \phi^* (x) \phi(y) \rangle = \frac{1}{2} (\langle Re \phi (x) Re \phi(y) \rangle + \langle Im \phi (x) Im \phi(y) \rangle)$ where $\langle Re \phi (x) Re \phi(y) \rangle=\langle Im \phi (x) Im \phi(y) \rangle$ = propagator of a real scalar field. This little note is for the purpose of avoiding misconceptions concerning Eq.~\ref{SD} where, after using non-relativistic concepts, some people mistakenly thought that the contribution from $\phi^0$ vanishes!
\end{thebibliography}
\end{document}